# Biologically Enhanced Energy and Carbon Cycling on Titan?

Dirk Schulze-Makuch[1] and David H. Grinspoon[2]

[1] Dept. of Geological Sciences, Washington State Universty, Pullman, WA 99164,

[2]Dept. of Space Studies, Southwest Research Institute, Boulder, Colorado

**With the Cassini-Huygens Mission in orbit around Saturn, the large moon Titan, with its reducing atmosphere, rich organic chemistry, and heterogeneous surface, moves into the astrobiological spotlight. Environmental conditions on Titan and Earth were similar in many respects 4 billion years ago, the approximate time when life originated on Earth. Life may have originated on Titan during its warmer early history and then developed adaptation strategies to cope with the increasingly cold conditions. If organisms originated and persisted, metabolic strategies could exist that would provide sufficient energy for life to persist, even today. Metabolic reactions might include the catalytic hydrogenation of photochemically produced acetylene, or involve the recombination of radicals created in the atmosphere by UV radiation. Metabolic activity may even contribute to the apparent youth, smoothness, and high activity of Titan's surface via biothermal energy.**

Environmental conditions are generally thought to be conducive for life if it can be shown that (a) polymeric chemistry, (b) an energy source, and (c) a liquid solvent are present in appreciable quantities (1). Polymeric chemistry has not been confirmed yet for Titan but is most likely present given the complex carbon chemistry in Titan's atmosphere and on its surface. Abundant energy sources are present at least in the form of UV radiation and photochemistry, and probably endogenic geological activity. Water as a liquid solvent may be limited, but liquid mixtures of water and ammonia are likely(2), and the recent Cassini radar images suggesting the presence of a young surface and ongoing cryovolcanism (3, 4) point towards near-surface liquid reservoirs.

The presence of large methane and ethane reservoirs with traces of dissolved $N_2$ on Titan's surface have been suggested (5), and their role as possible life-supporting solvents has been hypothesized (6).

There are many apparent heterogeneities on Titan's surface (7). Surface features observed at infrared and radar wavelengths may be of volcanic, tectonic, sedimentological or meteorological origin (3, 8). A methane cycle may exist on Titan with some similarities to the hydrological cycle on Earth. Methane clouds have been detected at Titan recently (9, 10), and methane rain is consistent with modelling results (11, 12). Due to its higher specific gravity, solid acetylene could be present on the bottom of the ethane-methane reservoirs and available as an energy source for various chemical reactions involving a multitude of organic compounds. The lack of obvious impact features in early high-resolution Cassini imagery (3, 4, 8) and inferred youth of the surface imply the possibility of active resurfacing and possible burial or subduction mechanisms that could supply subsurface liquid reservoirs with acetylene and other photochemical products. Fortes (2) pointed out that given the extremely cold surface, some of the simplest prebiotic reactions on Titan would have half-lives on the order of $10^7$ years. Perhaps, however, reactions could be accelerated by catalysis or localized warming. Regions of geothermal activity have been projected to exist on Titan (13), and early Cassini results suggest the presence of a young surface with widespread cryovolcanism (3). The most likely energy sources providing heat to the surface would be water volcanism or meteorite impacts (14). Both mechanisms may have created episodes of aqueous chemistry in lakes on Titan's surface, perhaps lasting thousands of years before freezing over (14, 15). An especially promising environment for life would thus be a hot spring or geothermal area at the bottom of a hydrocarbon reservoir, or an area where volatile overheated compounds intersect with such a reservoir (Figure 1). This environment would not only provide a versatile suite of raw material for organic synthesis and some amount of molten water and ammonia for organic reactions, but also higher temperatures for reactions to occur more rapidly. If such a site included microenvironments with fluid/solid interfaces, zones of fluid accumulation or entrapment, and areas enriched with material that could act as a catalyst (e.g. zeolites, clay) it would be especially favourable for biology (16).

Primitive metabolism can be envisioned as a reaction or reaction sequence that yields free energy. Once produced, the energy is used by the organism to do work. The nature of the metabolic strategies employed depends largely upon the environmental constraints that affect an organism. For instance, anaerobic bacteria use metabolic pathways that do not depend on oxygen. Organisms usually adapt to environmental conditions by evolving in a manner that allows them to utilize available raw materials. Due to the presence of a variety of carbon compounds in large quantities, any metabolic reaction pathway of a chemoautotrophic organism on Titan would probably involve the reduction or oxidation of at least one carbon compound.

Given the environmental conditions on Titan, a reasonable energy yielding reaction for a metabolizing microbe is the catalytic hydrogenation of photochemically produced acetylene:

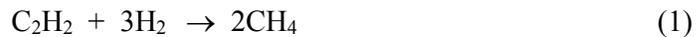

$$C_2H_2 + 3H_2 \rightarrow 2CH_4 \qquad\qquad (1)$$

The energy yield of this reaction is 107.7 kJ/mol ($\Delta$ G = -107.7 kJ/mol or –25.7 kcal/mol) under standard conditions, and about 100 kJ/mol under Titan surface conditions. Both acetylene and hydrogen are present in Titan's atmosphere at significant concentrations. Since acetylene is produced high in the stratosphere from solar UV radiation and then mostly condenses and falls to the surface, it provides a potential means of transferring high altitude solar UV energy to surface chemical reactions (15). Spectroscopic evidence suggests that the product of this reaction, methane, is found to be isotopically lighter than would be expected from theories of Titan's formation (17), and thus may hint toward microbial fractionation. The energy released by this reaction may be captured by the organism and used to drive an uphill reaction or it may be directly utilized to perform cellular work (6, 18).

Methane is unstable in Titan's atmosphere and is destroyed on a short time scale of about $10^7$ years by solar ultraviolet radiation (19). Without a constant re-supply, no significant amount of methane should be present in Titan's atmosphere. However, an atmospheric concentration of about 5 % is observed (15). Assuming that the re-supply is provided solely biologically via reaction (1) with no inorganic input, an energy of 2.2 x $10^{21}$ kJ is produced over a period of 10 million years to keep the observed methane concentration constant. The free energy of the anaerobic formation of one unit carbon formula weight (UCFW = 24 g/mol) for the yeast, *Saccharomyces cerevisiae*, has been determined to be 76.89 kJ (20), or 3.2 kJ/g. Assuming this energy demand and a generation time of 1 year, a biomass of $6.8 \times 10^{13}$ g at an energy conversion efficiency of 100 % would be required to keep the currently measured methane concentration at a constant level. Generation times for putative Titan organisms are assumed here to be much longer than for "average" terrestrial organisms, because of the slower kinetics in a colder environment. If the organisms were comparable in size to typical terrestrial microorganisms with a dry mass of $2 \times 10^{-14}$ g (21) and were envisioned to homogeneously populate the upper 1 m of the surface of Titan, the biomass density would be about $4.1 \times 10^{13}$ microbes per cubic meter, a typical density for slightly nutrient deprived environments on Earth. If there are significant inorganic sources of methane, the calculated biomass density would decrease accordingly.

On the other hand, the miniaturization of cellular life in water on Earth may be a misleading model for life in a non-aqueous environment (22). In an extremely cold, hydrophobic (but liquid) environment surface/volume ratio considerations may be less constraining than at higher temperatures in polar solvents. Thus, life on Titan could involve huge (by Earth standards) and very slowly metabolizing cells, in which case biomass densities would be higher than calculated above.

Other metabolic pathways are possible as well. An intriguing possibility is radical reactions as a basis for metabolism. Raulin (23) suggested that Titan's stratosphere is an active site of complex carbon and nitrogen radical chemistry. Thus, energy-yielding reactions could be based on chemistry involving radicals. For, example, a chemoautotrophic organism may use the following reactions

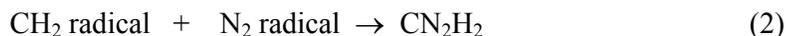

$$CH_2 \text{ radical} + N_2 \text{ radical} \rightarrow CN_2H_2 \qquad (2)$$

or

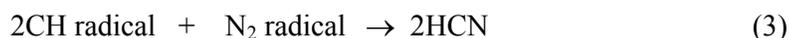

$$2CH \text{ radical} + N_2 \text{ radical} \rightarrow 2HCN \qquad (3)$$

Reactions (2) and (3) produce a high yield of energy and may take place in the atmosphere or at the surface of Titan. All reactants have been detected in Titan's environment based on data from Voyager 1 (24). On Earth, the energy-rich reactions involving radicals are very difficult to control and would cause internal damage to any organism. On Titan, however, at surface temperatures of less than 100 K, these reactions may proceed at a reasonable pace, and may constitute a feasible energy-yielding reaction for a metabolic pathway. The radical reactions (2) and (3) have the interesting side effect of producing the biologically important compounds cyanamide and hydrocyanic acid, respectively (6, 16).

The recent detection of the apparent youth and smoothness of Titan's surface and the likely evidence for active cryovolcanism, is puzzling given the probable lack of endogenic activity within Titan. Titan has a low bulk density of 1.88 g cm$^{-3}$, which would imply that silicate substrate is quite rare at the surface. The relative lack of heavy elements may cause less radioactivity and with it less volcanism and less heat. Nevertheless, the geothermal heat flow on Titan has been estimated as 4mWm$^{-2}$(25), which would suggest the presence of some volcanic activity, but may not be sufficient to account for the surface smoothness as observed. An intriguing possibility is that biological heating could contribute to the surface smoothness. Glacial melting from biothermal energy released by algal metabolism has been reported previously (26) as has been the influence of marine microorganisms on the melting of Arctic pack ice (27). Microbial colonization occurs in cryoconite holes on glaciers (28) and in basal glacial melt waters, some of which are known to release methane as metabolic end-product (29, 30). However, the question is how significant that effect is, or can be. In principle, using the suggested metabolic reaction (1) and applying it to Titan surface conditions a substantial amount of heating could be provided by biological means:

Given the Titan surface temperature of 95 K and the eutectic melting point of an ammonia-water mixture at 175 K, the energy that needs to be expended for melting is calculated by adding the heat capacity of the water-ammonia mixture to the enthalpy, resulting in approximately 11.5 kJ/mol. Of course, any putative organism would need some (really most) of the energy for its own metabolism. The energy used by the yeast organism *Saccharomyces cerevisiae* is 76.89 kJ/mol (see above), which compares to an estimated energy gain of 100 kJ/mol for reaction (1) under Titan's environmental conditions. Further, assuming again that 2.2 x 10$^{21}$ kJ of biological energy is produced over a period of 10 million years to keep the observed methane concentration constant with no inorganic input, this would require 2.2 x 10$^{14}$ kJ/year. Most of this energy will go toward the metabolism of the organism (~ 76.8 % assuming an organism with the same energy need as *Saccharomyces*), but the remaining energy (5.2 x 10$^{13}$ kJ/year) would in principle be available to heat the surrounding environment. Since 11.5 kJ is needed to melt 1 mol of water-ammonia rock, there is enough to melt 4.5 x 10$^{12}$ mol of water-ammonia ice in one melting event per year. This would be sufficient to melt 7.9 x $10^{10}$ kg of water and ammonia ice once per year.

These calculations indicate that biothermal melting would be possible. Given the low temperatures, the biological effect on Titan, if it exists, should be larger than on Earth. In conditions where the ability to sustain liquid microenvironments is a key limitation on survival, then adaptive pressures could lead to a larger percentage of the free energy of exothermic metabolic reactions going towards heating the immediate environments of organisms living close to the freezing point. On the other hand, much energy has to be expended to reach the liquid state. If volcanic activity or other energy sources are present and significant, it would increase the chances for life on Titan by elevating temperatures and providing potentially habitable geothermal areas and gases that could be used for metabolism. Any liquid water-ammonia mixture is lighter than the surrounding ice and will float if produced at depth.

Given the current sample size of one biosphere upon which astrobiologists must base their theories and speculations, our ideas about life elsewhere must remain fluid and not too heavily based upon the specific metabolisms, strategies and structures of terrestrial organisms. The basic requirements of life, as they are understood today, are all present on Titan, including organic molecules, energy sources and liquid media. If surface or subsurface organisms are able to take advantage of upper atmospheric photochemistry, through the continuous downward transport of high energy compounds such as acetylene, they would have vast energy reserves at their disposal which could be used, in part, to maintain the liquid environments conducive to life.

Acknowledgements

David Grinspoon's work was in part supported by grants from the NASA Exobiology

Research Program and the NSF Program of Research into Life in Extreme

Environments.

Figure 1. Environmental conditions at Titan, schematic. Acetylene and radicals are produced by photochemical reactions in the atmosphere. Due to its high specific gravity acetylene will sink to Titan's surface and to the bottom of a hydrocarbon reservoir, where it can be used by putative organisms for metabolic reactions (insert). The metabolic end-product methane rises to the atmosphere and is detected to be isotopically lighter than predicted by Titan formation theories.

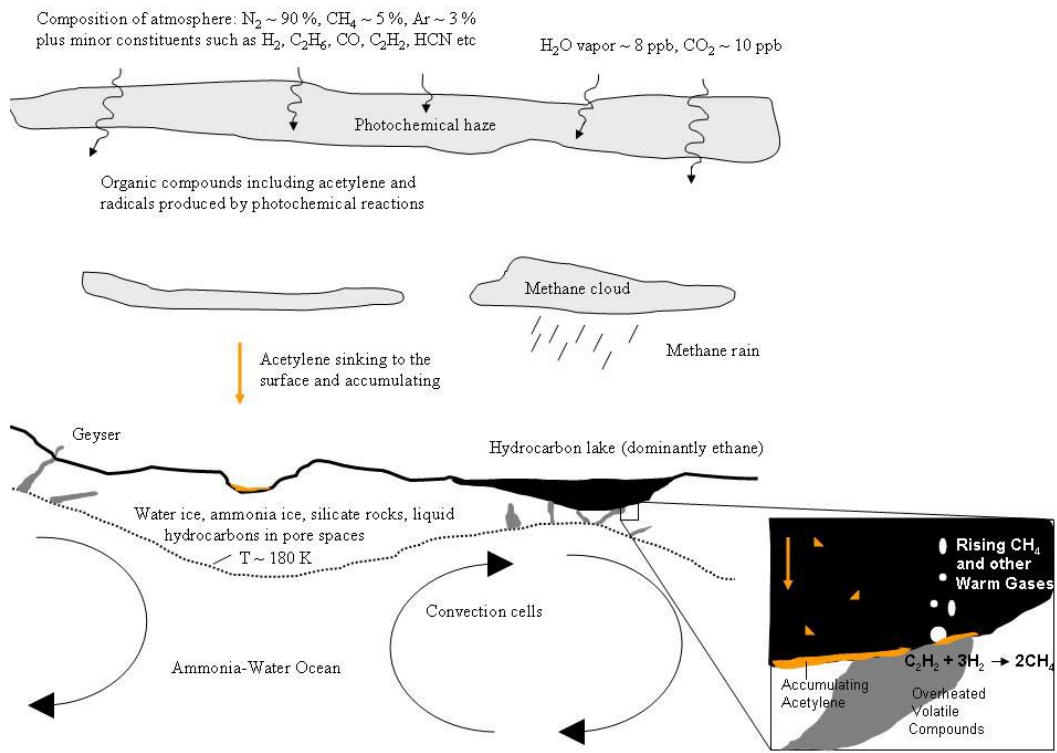